\documentclass[%
reprint,
nofootinbib,
amsmath,amssymb,
aps,
]{revtex4-2}

\pdfoutput=1
\usepackage{graphicx}%
\usepackage{dcolumn}%
\usepackage{bm}%
\usepackage[utf8]{inputenc}
\usepackage{mathrsfs}
\usepackage{color}
\usepackage{graphicx}
\usepackage{epstopdf}
\usepackage{amsmath}
\usepackage{amsfonts}
\usepackage{amssymb}
\usepackage{mathrsfs}
\usepackage{feynmp}
\usepackage{slashed}
\usepackage{latexsym}
\usepackage{verbatim}
\usepackage{subfigure}
\usepackage{longtable}
\usepackage[hyperfootnotes=false]{hyperref}
\usepackage{braket}
\usepackage{multirow}
\usepackage{textgreek}
\usepackage{comment}
\usepackage{xcolor}
\usepackage[normalem]{ulem}

\hypersetup{
	breaklinks=true,
	pdfusetitle=true
}

\newcommand{\bs}[1]{\boldsymbol{#1}}

\newcommand{\RNum}[1]{\uppercase\expandafter{\romannumeral #1\relax}}
\begin{document}
	
	\preprint{APS/123-QED}
	
	\title{A Possible Solution to the Difficulty in the Interpretation of Deuteron Compositeness}
	
	\author{Zanpeng Yin}
	\email{yin.z.aa@m.titech.ac.jp}
	\affiliation{Department of Physics, Tokyo Institute of Technology.}
	\author{Daisuke Jido}
	\email{jido@th.phys.titech.ac.jp}
	\affiliation{Department of Physics, Tokyo Institute of Technology.}

	\date{\today}%
	
\begin{abstract}
We study the theoretical structure of compositeness with explicit energy dependence, and find a possible explanation for the difficulty in the interpretation of compositeness of deuteron. Compositeness of deuteron is calculated as larger than one in many methods like weak-binding limit. Even though it is widely assumed that the energy dependence in interaction always comes from other states, which we call surjective interpretation, we find that the outcome of deuteron may suggest a violation of surjective interpretation. We directly perform numerical and perturbative calculations of deuteron compositeness. It is concluded that if the energy dependent part of interaction contributes to attraction, compositeness is likely to be enhanced from unity. We discuss the indications of this outcome and the model dependence of compositeness. We propose a straightforward extension and a thorough revise on the formalism of compositeness with field theory considerations.
		\begin{description}
			\item[Keywords]
			Hadron Physics, Compositeness, Deuteron, Effective Theory
		\end{description}
\end{abstract}
	
	\maketitle
	
\section{Introduction}
Hadrons are composed of quarks and gluons which are governed by Quantum Chromodynamics (QCD). Normal hadrons are made up by a pair of a quark and an antiquark like mesons or three quarks like baryons, and these hadrons form the basis of low energy effective theory for hadrons. However, some outliers exist outside of such categories, and they are named exotic hadrons. While more hadronic composite states are allowed in QCD, we can only observe several of such states, which makes the research of exotic hadrons more important. Some currently observed exotic hadrons include $\Lambda$(1405)\cite{Hyodo2016},  X(3872)\cite{PhysRevLett.91.262001}, Y(4260)\cite{PhysRevLett.96.162003}, etc. The structure of exotic hadrons are proposed to be elementary or composite. From the aspect of hadron physics, elementary states are particle states that cannot be expressed by other hadrons, and composite states are composed by other hadrons with a larger size than normal hadrons. However, quark radial excitation, quark pair creation, and meson creation are all on similar energy scales, so these excitation modes may compete with each other in hadron structure. A hadron state may, as a result, be a mixture among those states. To better understand the hadron structure, separating these contributions from different origins is important.\par 
Compositeness is a quantity proposed by S. Weinberg in Refs.\cite{Weinberg1963,Weinberg1965} to distinguish whether deuteron is constructed by nucleons or a distinct elemental particle composed of, for example, six quarks. Compositeness is defined by the proportion of a state composed of other more elementary particles, in this case the proportion of deuteron being composed of nucleons. This procedure can be seen as a mixture of elementary states and composite states to a physical state. S. Weinberg formalized his calculation based on nucleon scattering in the low energy regime. With current experimental values, this method resulted in a compositeness larger than unity, which means a negative elementariness. However, it is not a reasonable outcome because elementariness, same as compositeness, is defined to be a proportion, and a proportion should not be negative.

In the past century, works are done on the properties and calculations of compositeness mostly in a Quantum Field Theory (QFT) scheme \cite{Hite1974, Salam1962}, and a modern QFT version of compositeness is given in Ref.\cite{Oller2018}. Recently, compositeness has been developed by T. Sekihara et al. and utilized in the description of various particles \cite{Sekihara2015,Kamiya2017, Sekihara2017, Sekihara2021}. They have derived the formulae of compositeness and elementariness based on Schr\"odinger equation and extended them for quantities obtained in Lippmann-Schwinger scattering equation by interpreting that any energy dependence in the interaction between hadrons as outcomes of elementary states and other scattering channels. This extension is practically important because one is able to use scattering experimental data to fix the hadronic interaction and evaluate compositeness and elementariness based on experimental observables. By taking weak-binding limit, the calculation of S. Weinberg can be recovered. The idea of weak-binding limit has been developed in Refs.\cite{Kamiya2017,Hyodo2013,Kamiya2016,Albaladejo2022,Kinugawa2022}. The importance of weak-binding limit lies in the fact that this method is almost model independent, i.e. compositeness can be calculated from physical observables only. Various theoretical works on the interpretation and calculation methods of compositeness are performed in Refs.\cite{Li2022,Song2022,Guo2016, Hyodo2012-2,Kamiya2017, Bruns2019}.  On the other hand, model calculation without using weak-binding limit has been performed in Refs.\cite{Sekihara2015,Hyodo2012, Ahmed2020,Aceti2014}. Numerical calculation of compositeness utilizing Lattice QCD calculation has been carried out in Refs.\cite{Bali2017,Torres2015}\par 
Despite the wide discussion and usage of compositeness, few has been done to confirm its soundness. The compositeness of deuteron is still larger than one, or with negative elementariness, in the current formalism. On the other hand, we expect that deuteron is at least almost composite, which means compositeness is almost unity but smaller. In this paper, we would aim at filling the gap between calculation and expectation by testing the soundness of compositeness as a physical quantity from a hadron physics point of view .\par
In Section \ref{section2}, we will first review the modern formalism of T. Sekihara et. al. with generalization into energy dependent potential, and the positivity of elementariness is derived from a theoretical viewpoint. Under surjective interpretation, where the energy dependent part of the interaction is always interpreted to come from other states, T. Sekihara's formalism can be recovered. On the other hand in Section \ref{section3}, we find that binding energy higher than that indicated by the scattering length is likely to result in an unphysical negative elementariness. In Section \ref{section4}, we discuss the reason and indication of this difficulty and proposed some alternative approaches for the phenomenological calculation of compositeness. Section \ref{section:conclusion} is devoted to conclusion of this work.

\section{Model and Formalism}
\label{section2}
In this section, we review the formalism of compositeness with development in energy dependent potential. To do this, we first define the model space and compositeness in a theoretical point of view. Then we demonstrate the calculation of compositeness from Schrodinger equation which can be connected to the scattering theory with surjective interpretation which will be defined later \cite{Sekihara2015}.
\subsection{Model and Definition of Compositeness}

Let us consider a scenario which is composed by scattering states in several channels $\ket{q_j}$ and a number of elementary states $\ket{\psi_a}$. Each scattering state $\ket{q_j}$ is composed of two free particles with mass $m_j$ and $M_j$ propagating with momentum $q_j$ in the center-of-mass(CM) frame. We use the subscript $j$ to indicate different channels. The elementary states are one-body states and cannot be described by the interaction of the scattering states.

The full Hamiltonian $\bs{H}$ for this system consists of two parts,
\begin{equation}
	\bs{H}=\bs{H}_0+\bs{V}
\end{equation}
where $\bs{H}_0$ is the Hamiltonian when no interaction exists and $\bs{V}$ is the interaction between eigenstates of this free system, which may depend on the energy of the system $E$ explicitly.

We will first define the eigenstates in the free system as
\begin{align}
	\bs{H}_0\ket{q_j}&=E_{q_j}\ket{q_j},\notag\\
	\bs{H}_0\ket{\psi_a}&=M_{B_a}\ket{\psi_a},
\end{align}
where
\begin{align}
	E_{q_j}&=M_{th,j}+\frac{q_j^2}{2\mu_j},
	\label{onshell}
\end{align}
with the threshold energy $M_{th,j} = m_j+M_j$ and the reduced mass $\mu = \frac{m_j M_j}{m_j+M_j}$.  We also explicitly define one body elementary states $\ket{\psi_a}$ with no requirement on the origin of these states. The normalization writes as follows,
\begin{align}
	\braket{{q'}_j|q_k}&=(2\pi)^3\delta(q'-q)\delta_{jk},\notag\\
	\braket{\psi_a|\psi_b}&=\delta_{ab},\notag\\
	\braket{\psi_a|q_k}&=0.
\end{align}
We assume that these states form a complete orthonormal basis, which we call model space, as
\begin{equation}
	1=\sum_a\ket{\psi_a}\bra{\psi_a}+\sum_j\int\dfrac{d^3q}{(2\pi)^3}\ket{q_j}\bra{q_j} \label{complete} .
\end{equation}

By including the interaction term $\bs{V}$, it generates a physical bound state $\ket{\psi}$ with the eigenenergy of the full Hamiltonian $\bs{H}$ or mass $M_B$ as
\begin{equation}
	\bs{H}\ket{\psi}=M_B\ket{\psi},
	\label{M_B}
\end{equation}
where we normalize the bound state $\ket{\psi}$ as unity:
\begin{equation}
	\braket{\psi|\psi}=1.
	\label{renormalize}
\end{equation}

From here, we define compositeness $X$ and elementariness $Z$ as
\begin{align}
	X_j&=\int\dfrac{d^3q_j}{(2\pi)^3}\braket{\psi|q_j}\braket{q_j|\psi},\notag\\
	Z_a&=\braket{\psi|\psi_a}\braket{\psi_a|\psi}.
\end{align}
By defining the wave function of the bound state as $\tilde{\psi}_j(q)=\braket{q_j|\psi}$ in momentum space, the compositeness can also be written as
\begin{equation}
	X_j=\int\dfrac{d^3q_j}{(2\pi)^3}\tilde{\psi}_j^*(q)\tilde{\psi}_j(q) .
\end{equation}
The compositeness represents the proportion of the physical state composed by a scattering channel, and elementariness is the proportion of that composed by the elementary state. From the assumption of the complete Hilbert space in Eq.\eqref{complete}, and the definition of the normalization factor in Eq.\eqref{renormalize}, we can achieve a normalization of elementariness and compositeness as
\begin{equation}
	1=\sum_aZ_a+\sum_jX_j,
\end{equation}
which is in accordance with our intuition that the sum of probability of all possible events ought to be one.

For later convenience, we assume a separable form in the momentum space:
\begin{align}
	\braket{{q}_j|\boldsymbol{V}|{q'}_k}&=v_{jk}(E)f_j(q^2)f_k(q'^2)\notag\\
	\braket{\psi_a|\bs{V}|{q'}_j}&=g_{(0)a,j}f_j(q'^2)\notag\\
	\braket{\psi_a|\bs{V}|\psi_b}&=0,
	\label{V elements}
\end{align}
in which $f(q^2)$ is a form factor, and coupling $g_{(0)a,j}$ is assumed to be constant. We can choose an appropriate phase so that $v_{jk}$ and $g_{(0)a,j}$ are real. We assume that the interaction $v_{jk}$ may depend on an external energy $E$, which we do not specify here.

\subsection{Compositeness from Schrodinger Equation: Single Channel}
\label{subsection.singlechannel}
In this section, we will mostly follow the formalism developed in Ref.\cite{Sekihara2015} and pay special attention to energy dependent potentials. We will first discuss the compositeness of a bound state in single channel case. We assume one scattering channel $\ket{q}$ and one elementary state $\ket{\psi_0}$ in the free theory.

The wave function $\tilde{\psi}(q)=\braket{q|\psi}$ for the bound state $\ket{\psi}$ is represented by the coupled-channel Schrodinger equations,
\begin{align}
	\braket{q|\bs{H}|\psi} &= f(q^2)\int \dfrac{d^3q'}{(2\pi)^3} v(E) f(q'^2)\tilde{\psi}(q') \notag\\
	&+E_q\tilde{\psi}(q)+g_{(0)}f(q^2)\braket{\psi_0|\psi}= M_B\tilde{\psi}(q),
	\label{comp2_eq1}
\end{align}
\begin{align}
	\braket{\psi_0|\bs{H}|\psi}=M_{B_0}\braket{\psi_0|\psi}+g_{(0)}\int &\dfrac{d^3q'}{(2\pi)^3}f(q'^2)\tilde{\psi}(q') \notag\\
	& =  M_B\braket{\psi_0|\psi}.
	\label{comp2_eq2}
\end{align}
These equations are obtained by taking the matrix elements of Eq.\eqref{M_B}, inserting the complete set Eq.\eqref{complete} and using the matrix elements of $\bm{V}$ from Eq.\eqref{V elements}.

Through Feshbach method \cite{Feshbach1958}, we can reduce the influence of the elementary channel as a part of the interaction in the scattering channel. We further assume the external energy to be the bound state energy, resulting in an effective Schrodinger equation,
\begin{align}
	E_q\tilde{\psi}(q)+f(q^2)v^{\text{int}}(M_B)\int \dfrac{d^3q'}{(2\pi)^3}&f(q'^2)\tilde{\psi}(q')\notag\\
	&= M_B \tilde{\psi}_j(q),
	\label{sch_eq}
\end{align}
with an integrated potential defined as
\begin{equation}
	v^{\text{int}}(E)=v(E)+\dfrac{(g_{(0)})^2}{E-M_{B_0}}.
	\label{veff1}
\end{equation}
The integrated potential $v^{\text{int}}(E)$ contains the effect of the elementary state acting on scattering states, in addition of the interaction $v(E)$ inside the scattering states.\footnote{In Ref.\cite{Sekihara2015}, $v^{\text{int}}(E)$ is called effective potential, while we call it integrated potential to distinguish it from the effective potential originated from an effective theory.}

Formally, the solution of Eq.\eqref{sch_eq} can be written as
\begin{align}
	\tilde{\psi}(q) = - \dfrac{cf(q^2)}{E_{q}-M_B}
	\label{comp2_wave},
\end{align}
with
\begin{align}
	c=v^{\text{int}}(M_B)\int \dfrac{d^3q'}{(2\pi)^3}f(q'^2)\tilde{\psi}(q') .
	\label{comp2_c}
\end{align}
We need to note this set of equations is equivalent to Schordinger equation \eqref{sch_eq} and still yet to be solved.

In order to have a non-trivial solution, we require a non-zero $c$. By inserting Eq.\eqref{comp2_wave} into Eq.\eqref{comp2_c}, we obtain
\begin{equation}
	1-v^{\text{int}}(M_B)\int \dfrac{d^3q'}{(2\pi)^3}\dfrac{[f(q'^2)]^2}{M_B-E_{q'}} = 0.
\end{equation}
By introducing a function $G$ as
\begin{equation}
	G(E) = \int \dfrac{d^3q}{(2\pi)^3}\dfrac{[f(q^2)]^2}{E-E_{q}},
	\label{comp2_G}
\end{equation}
which turns out to be the Green's function that represents the propagator in the free system. We will be able to arrive at the bound state condition
\begin{equation}
	1=v^{\text{int}}(M_B)G(M_B),
\end{equation}
which will also be seen in Lippmann-Schwinger equation later.

By utilizing Eq.\eqref{comp2_wave} and Eq.\eqref{comp2_G}, the compositeness is expressed as
\begin{align}
	X &= \int \dfrac{d^3q}{(2\pi)^3}|\tilde{\psi}(q)|^2=-|c|^2[\dfrac{dG}{dE}]_{E=M_B}.
	\label{X_calculate}
\end{align}
Similarly, we also express the elementariness of a bound state, which is
\begin{align}
	Z&=\braket{\psi|\psi_0}\braket{\psi_0|\psi} \notag\\&= |c|^2 G(M_B)\dfrac{g_{(0)}g_{(0)}}{(M_B-M_{B_0})^2}G(M_B)\notag\\
	&= - |c|^2 [G\dfrac{d(v^{\text{int}}-v)}{dE}G]_{E=M_B},
	\label{Z}
\end{align}
where we have inserted Eq.\eqref{comp2_wave} into Eq.\eqref{comp2_eq2}. We note that, if $v$ depends on the energy $E$, we should subtract the contribution of $v(E)$ from the elementariness, because the elementariness is defined by the attribution of the elementary state. Thanks to the completeness of the Hilbert space and normalization of $\braket{\psi|\psi}=1$, we will have
\begin{equation}
	- |c|^2 [\dfrac{dG}{dE}+G\dfrac{d(v^{\text{int}}-v)}{dE}G]_{E=M_B}= X+Z = 1.
	\label{identity}
\end{equation}
\subsection{Compositeness from Schrodinger Equation: Multiple Channels}
We have discussed the formalism of compositeness for a single channel of interest. This formalism can be extended to multi-channels, and also for several elementary states straightforwardly as is described in Ref.\cite{Sekihara2015} for constant interactions. We will refrain from giving the full formalism, and only the outcome will be stated.

For a bound state with $E=M_B$, we have
\begin{align}
	X_j &= \int \dfrac{d^3q}{(2\pi)^3}|\tilde{\psi}_j(q)|^2=-|c_j|^2[\dfrac{dG_j}{dE}]_{E=M_B},\\
	Z_a&=\braket{\psi|\psi_a}\braket{\psi_a|\psi}\notag\\ &=\sum_{j,k} c_j c_k G_j(M_B)\dfrac{g_{(0)a,j}g_{(0)a,k}}{(M_B-M_{B_a})^2}G_k(M_B),\\
	Z &= \sum Z _a = - \sum_{j,k}c_j c_k [G_j\dfrac{d(v_{jk}^{\text{int}}-v_{jk})}{dE}G_k]_{E=M_B},
\end{align}
in which the subscripts notate the entries of the matrix that is used to notate multichannels. The wave function $\tilde{\psi}_j(q)$ is defined by $\tilde{\psi}_j(q)=\braket{q_j|\psi}$, and $c_j$ is defined in a similar way as Eq.\eqref{comp2_c}, $G_j$ is the loop function in channel $j$, $X_j$ is the compositeness in $j$ channel, $Z_a$ is the elementariness as a result of the existence of state $a$, and $v_{jk}^{\text{int}}$ is the integrated potential between channels $j$ and $k$ as
\begin{equation}
	v_{jk}^{\text{int}}=v_{jk}(E) + \sum_a \frac{g_{(0)a,j} g_{(0)a,k}}{E-M_{B_a}}.
\end{equation}

The compositeness and the total elementariness can both be found in a similar form as in the single channel case. Since the compositeness is related to the completeness of the model space, we will have a normalization, which writes
\begin{equation}
	- \sum_{j,k}c_j c_k [\delta_{jk}\dfrac{dG_j}{dE}+G_j\dfrac{d(v_{jk}^{\text{int}}-v_{jk})}{dE}G_k]_{E=M_B} = 1.
\end{equation}
We need to note that compositeness, as well as elementariness, are highly influenced by the way we define the dynamics.

We can include an explicit channel $N$ into integrated potential through Feshbach method and make them implicit\cite{Sekihara2015,Feshbach1958}, by defining
\begin{equation}
	w_{jk} = v^{\text{int}}_{jk} + \sum_{j,k\neq N} v^{\text{int}}_{jN}
	\dfrac{G_N(E)}{1-v^{\text{int}}_{NN} G_N(E)}v^{\text{int}}_{Nk},
	\label{veff2}
\end{equation}
where $N$ indicates the channel to be reduced, and the contribution of this channel is absorbed into the integated interaction $w_{jk}$ of this channel. We call such channels reduced channels.
Because the contribution of the reduced channels gives additional energy dependence to the integrated interaction, the reduced channels can be another source of the elementariness, even though the contribution of the reduced channels is originally counted as the compositeness before the channel reduction.

This reduction can be performed on multiple channels. It means that we are able to send the influence of explicit channels into the integrated interaction, and the original compositeness of the reduced channels will appear as elementariness. We can find that this reduction is similar to what we do to the bare states and absorb its contribution into $v^{\text{int}}$, in which the explicit elementary state becomes implicit. Effectively, we reduce the model space into its subspace which will be actually solved, then the integrated interaction has energy dependence and the elementariness is evaluated by such energy dependence.

\subsection{Scattering Theory and Surjective Interpretation}
From scattering theory, the generation of a bound state can be described through the scattering process of its components:
\begin{align}
	\bs{T} = \bs{V} + \bs{VG}\bs{T},
\end{align}
in which $\bs{G}$ is the Green's function expressed by the propagator defined in the free Hamiltonian $\bs{H}_0$. In our definition the $T$-matrix is related with the scattering amplitude by
$\bs{S} = 1+ i \frac{q \mu}{\pi}\bs{T}$, in which $q$ is the momentum in the CM frame, and $\mu$ is the reduced mass of the system.

Here we consider a single two-body channel scattering state and a one-body elementary state for simplicity. The generalization to multi-channels is straightforward as done in the previous sections.

The Green's function $\bs{G}$ takes a form of $\frac{1}{E-\bs{H}_0}$. By inserting the complete set we have
\begin{align}
	\bs{T} &= \bs{V}^{\text{int}}(E) + \bs{V}^{\text{int}}(E)\ket{q}\dfrac{1}{E-E_{q}}\bra{q}\bs{T},\\
	\bs{V}^{\text{int}}&= \bs{V} + \bs{V}\ket{\psi_0}\dfrac{1}{E-M_{B_0}}\bra{\psi_0}\bs{V}.
\end{align}
As a result, the matrix element of the operator $T$ for the scattering state $\ket{q_j}$ writes as
\begin{equation}
	t(E)=\dfrac{1}{1-v^{\text{int}}(E)G(E)}v^{\text{int}}(E),
	\label{LSeq_eff}
\end{equation}
where $v^{\text{int}}$ and $G$ are the matrix elements of the operators $\bs{V}^{int}$ and $\bs{G}$ for the scattering states, respectively. Since we ensure in Eq.\eqref{LSeq_eff} that a pole, i.e. a bound state exists as $v^{\text{int}}(M_B)G(M_B)=1$, we are able to get a residue of this pole, as
\begin{align}
	g^2&=\lim_{E\rightarrow M_{B}}(E-M_B)t(E)\notag\\
	&=- \dfrac{1}{[\frac{dG}{dE} + \frac{1}{(v^{\text{int}})^2 }  \frac{dv^{\text{int}}}{dE}]}.
	\label{g2}
\end{align}

Equation \eqref{g2} is equilavent to

\begin{equation}
	- |g|^2 [\dfrac{dG}{dE}+G\dfrac{dv^{\text{int}}}{dE}G]_{E=M_B}= 1.
	\label{identity_g}
\end{equation}
The residue ought to be positive as it is related to the renormalization factor of the bound state, and the absolute value is taken to keep consistency. We can find that Eq.\eqref{identity_g} shares a similar but not exactly identical form with Eq.\eqref{identity}. This difference originates the explicit $E$ dependence of $v$ in $v^{int}$. $c$ and $g$ are connected as

\begin{equation}
	|g|^2 = \frac{|c|^2}{1+|c|^2[G\frac{dv}{dE}G]_{E=M_B}}.
	\label{gc_renormalization}
\end{equation}
If we assume a constant $v$, we conclude $c=g$ from Eq.\eqref{gc_renormalization}.

Comparing Eq.\eqref{gc_renormalization} with Eq.\eqref{identity} and interpreting that all of the energy dependences of the integrated interaction stem from elementariness, Sekihara et al.\ have proposed in Ref.\cite{Sekihara2015} that compositeness and elementariness can be calculated in Lippmann-Schwinger formalism as

\begin{align}
	\breve{X} &=-|g|^2[\dfrac{dG}{dE}]_{E=M_B}, \label{X_g}\\
	\breve{Z} & = - |g|^2 [G \frac{d v^{\rm int}}{dE} G]_{E=M_B}.\label{Z_g}
\end{align}
Evaluating compositeness and elementariness by Eqs. \eqref{X_g} and \eqref{Z_g} relies on surjective interpretation, which is defined as the scenario that the energy dependence in interaction always comes from other states. To avoid confusion, we use different notations to distinguish from the original definition of compositeness and elementariness from Schr\"odinger formalism shown in Eqs.\eqref{X_calculate} and \eqref{Z}. When the interaction $v$ is a constant, Eqs. \eqref{X_g} and \eqref{Z_g} are equivalent to \eqref{X_calculate} and \eqref{Z}, as T. Sekihara et al. proved in Ref.\cite{Sekihara2015}, but a constant $v$ means that the surjective interpretation needs to be satisfied.

\subsection{Positivity of Elementariness \texorpdfstring{$Z$}{Z}}
\label{properties of compositeness}
In this subsection, we will prove the positivity of elementariness if the energy dependence of the integrated interaction stems from physical states. This proof is partly based on the nature of the Hilbert space as an inner product space. We will be assuming $E=M_B$, as only the bound state is considered.
\subsubsection{One-body State}
As can be seen in Eq.\eqref{veff1} and Eq.\eqref{Z}, the contribution of any elementary states on elementariness ought to be positive. It is because we have included real elementary states, and thus such states are a part of our Hilbert space, where the norm ought to be positive.
\subsubsection{Two-body State}

We have introduced the integrated potential coming from a scattering channel in Eq.\eqref{veff2}.  Considering channel $N$ to be reduced into channel 0, where interactions $v_{0N}$ and $v_{NN}$ are constant, we have 

\begin{align}
	\frac{\partial (w_{00} - v_{00})}{\partial E} &= v^2_{0N}
	\frac{\partial}{\partial E}  \dfrac{G_N(E)}{1-v_{NN} G_N(E)}
	\notag\\
	&=v^2_{0N} \frac{\partial G_N}{\partial E} \frac{1}{(1-v_{NN}G_N)^2} <0,
\end{align}
where we have used the conclusion that $\frac{\partial G_N}{\partial E}<0$ when the energy is below the threshold. It is valid because the loop function is limited by optical theorem. We are actually able to derive this sign from the mathematical form of loop function,
\begin{align}
	G(E)&=-\int_{s_+}^{\infty} ds' \frac{\rho(s')}{(s'-E^2)}, \label{G_general}\\
	G'(E)&=-\int_{s_+}^{\infty} ds' \frac{\rho(s')}{(s'-E^2)^2}(2E)<0, \label{G_general_derivative}
\end{align}
in which $\rho$ represents the phase space, and $G$ is a real function below the threshold. \footnote{
One may consider subtracted dispersion relations instead of Eq.\eqref{G_general}. The subtraction terms will  introduce energy dependence which is not constrained by unitarity. Nevertheless, for practical use for hadronic phenomenology, Eq.\eqref{G_general_derivative} is still satisfied, because the loop function for two body scattering is converge with once subtracted dispersion relation, which introduces only a subtraction constant and it does not contribute to the derivative. 
}

From Eq.\eqref{veff2}, we always have positive elementariness when including a two-body channel with constant interactions, which is similar to including an explicit state.

\subsubsection{Three-body State}

The same statement can be extended to 3-body states. The only difference from a 2-body scattering state and a 3-body scattering state can be found only in the loop function. Let us take the case of deuteron, and define the 3-body state as a $NN\pi$ state, which represents the pion exchange procedure. Note that the mass is excluded from the energy in this subsection for simplicity. If we define the reduced mass according to Ref.\cite{Thomas:1977ph} as
\begin{align}
	\mu_{NN}&=\frac{m_N}{2},\\
	\mu_{\pi}&=\frac{m_\pi (m_N+m_N)}{m_N+m_N+m_\pi}=\frac{2m_\pi m_N}{2m_N+m_\pi},
\end{align}
we will be able to express this 3-body loop function as
\begin{align}
	G&(E)=\int \frac{d^3 p_\pi}{(2\pi)^3} \frac{d^3 q_{NN}}{(2\pi)^3}
	\frac{1}{E+i\epsilon-\frac{p_\pi^2}{2\mu_\pi}-\frac{q_{NN}^2}{2\mu_{NN}}}\notag\\
	&=\int \frac{d^3 p_\pi}{(2\pi)^3}
	[-\frac{2\mu_{NN}\Lambda_{NN}}{2\pi^2}
	-\frac{2 i \mu_{NN}}{4\pi}\sqrt{2\mu_{NN}(E-\frac{p_\pi^2}{2\mu_\pi})}
	],
\end{align}
where $\Lambda_{NN}$ is the cutoff of two body loop function $G_{NN}$, and we utilize a loop function that is stated in Ref.\cite{BEANE2001} with a modification by adding a constant to adapt to some more proper calculations on renormalizatons. Since the loop function is constrained by optical theorem, taking this form for calculation will not lose generality.

We will not evaluate the loop function directly, and only look at the sign of $\frac{dG}{dE}$,

\begin{align}
	\frac{dG}{dE}=-\frac{ \mu_{NN}^{\frac{3}{2}}}{8\sqrt{2}\pi^2}\int_0^\infty dp_\pi p_\pi^2 \frac{1}{\sqrt{-E+\frac{p_\pi^2}{2\mu_\pi}}}.
\end{align}
In the area of interest, $E<0$, and thus $\frac{dG}{dE}<0$. We need to note that even there is a square root in the expression, the Riemann surface structure is not relavent. It is because the branch cut in scattering theory is usually taken from the threshold pointing to the positive direnction.

In this way, the elementariness generated by 1,2,3-body scattering states with constant interaction is proved to be positive. This proof is made based on the definition of elementariness \eqref{Z} and the sign of loop functions. This outcome can also be interpreted in a way that the energy dependence stems form the reduction of the Hilbert space and the completeness should be kept in the channel reduction. Further, when taking surjective interpretation, it is natural that we conclude that elementariness $\breve{Z}$ evaluated by Eq.\eqref{Z_g} is also positive.

\subsection{Discussion: Formalism of Compositeness}

We have reviewed the calculation method of compositeness and elementariness based on scattering theory. We also concluded that, if the energy dependence of the integrated interaction comes from the channel reduction, the elementariness $\breve{Z}$ evaluated by Eq.\eqref{Z_g} is expected to be positive. We have also clarified that the calculation through Eq.\eqref{X_g} and \eqref{Z_g} as described in Ref.\cite{Sekihara2015} relies on surjective interpretation, that is, the method works with a positive value of elementariness when the energy dependence comes soley from physical states. In later sections, however, we will see that Eqs.\eqref{X_g} and \eqref{Z_g} may provide a negative value of $\breve{Z}$ in certain situations, such as deuteron. 

To connect the compositeness and elementariness in the Schrodinger formalism and the Lippmann Schwinger formalism, one needs to pin down the origin of the energy dependence in the integrated interaction $v^\textrm{int}$. It is possible when one knows a priori that the intrinsic energy dependence is negligibly smaller than that of the other energy dependence. Alternatively, we can also pin down the intrinsic energy dependence by using effective theories of the scattering system, such as chiral effective theories. For this purpose, we need to introduce models to describe the integrated interaction and the evaluation of compositeness becomes more model dependent. 

We would like to emphasize that using the scattering equation in the calculation of compositeness and elementariness has a practical advantage, because scattering data are available for the determinations of the residue of the scattering amplitude at the bound state pole and the energy dependence of the integrated interaction, which allows us to evaluate compositeness and elementariness phenomenologically based on physical observables. This is one of the most important findings in Ref.\cite{Sekihara2015}.  As phenomenological investigations, one may parameterize the integrated interaction of the scattering system in interest by assuming minimal requirements, such as unitary and symmetries. With enough experimental data, these parameters can be determined by scattering observations, and evaluation of compositeness and elementariness in a less model dependent way becomes possible.

It is worth noting that in Ref.\cite{Sekihara2015}, the calculation based on surjective interpretation was applied to several of the candidates of exotic hadrons. With chiral coupled-channel scattering models, the authors were able to conclude that the higher pole of $\Lambda(1405)$ and $f_0(980)$ are dominated by the composite states of $\bar{K}N$ and $\bar{K}K$ respectively, while the vector mesons $\rho(770)$ and $K^*(892)$ are dominantly elementary.

We would like to note that the word energy-dependent may be unnatural in some research areas, as the full Hamiltonian should not have any energy dependence. However, it is a natural concequence for a phenomenological description with Feshbach Partitioning \cite{Feshbach1958}, and enables a straightforward connection between this paper and its predecessors. For completeness, we provide an alternative formalism with only velocity dependent but no explicit energy dependence in Appendix.\ref{append.0}. This alternative formalism assumes a form of $\bs{V}(q^2,q'^2,E(q^2))$ and is equivalent to an explicit energy dependent formalism $\bs{V}(q^2,q'^2,E)$ under an ansatz, which can be tested by the bound state condition.

To be clear, our definition of compositeness and elementariness are based on quantum mechanics, in which one requires clearly defined intermediate states when one inserts the complete set as intermediate states. This is formulated as old-fashioned perturbation theory (OFPT) \cite{Schwartz:2014sze}, in which every intermediate state is on-shell while energy conservation is violated at vertices. Thus, the energy of the intermediate state is well fixed and there is no room for the intrinsic energy dependence for the integrated interaction. QFT formulates perturbation theory differently, in which four-momentum is conserved at vertices and intermediates can be off-shell. Because the intermediates do not have definite energy, the integrated interaction can be energy dependent. Of course, it is known that both formulations provide equivalent results on the observations, but some off-shell behaviors can be different. 

\section{Application to Deuteron: Complication of Phenomenology}
\label{section3}
Even though Eqs. \eqref{X_g} and \eqref{Z_g} show promising indication that compositeness and elementariness can be obtained by using scattering observables, it has been shown that a numerical calculation of compositeness of deuteron is larger than one. When Weinberg proposed compositeness, he intended to answer the question whether deuteron is a composite particle or not. He presented a calculation framework under weak-binding limit and obtained 1.68 compositeness with physical observables such as scattering length and effective range \cite{Weinberg1963}\cite{Weinberg1965}. Since compositeness should be interpreted as a proportion, a compositeness larger than one is not natural, even though S. Weinberg concluded that deuteron is at least almost composite.

This section will be dedicated to dicussing the difference between $X$ and $\breve{X}$, $Z$ and $\breve{Z}$. We will demonstrate the calculation of the compositeness of deuteron based on Eq.\eqref{X_g}, a.k.a. $\breve{X}$, using simple models and see that the value of compositeness turns to be greater than one when we use observed $NN$ scattering properties and deuteron binding energy. We will also find out that the origin of the problems is in the calculation method \eqref{X_g} and \eqref{Z_g}. Further, we will show that it is not likely to be avoidable for the deuteron case. Finally we discuss phenomenological implication of the findings of this section.
\subsection{Simple Model Calculation}
Based on Eqs.\eqref{X_g} and \eqref{Z_g}, we perform the calculation of the compositeness of the deuteron by introducing an effective interaction, and it turns out that the value of deuteron compositeness becomes greater than unity. We use simple models for the integrated interaction by considering a low momentum expansion. The parameters of the integrated interaction are fixed by the observables such as the scattering length, the effective range and the deuteron binding energy. Besides, the calculation is based on a local and separable form of interaction. For more sophisticated calculations, we can use whole amplitudes of the $NN$ scattering at low energy, which are available in market.

The simplest model may be a constant interaction. We can fix the value of the constant interaction with the scattering length. This model can reproduce neither the deuteron binding energy nor effective range. Actually, this model underestimates the deuteron binding energy. The value of compositeness of this simple model is one by definition because the interaction has no energy dependence and the elementariness is zero.

In order to provide a more realistic deuteron binding energy, we need an attractive correction in the effective interaction of the S-wave. In the momentum expansion, we use $c_0 + c_1 q^2$ and the parameters are fixed by the scattering length and the effective range. The deuteron binding energy is calculated as 2.232 MeV, which is consistent with the observation. Because the $q^2$ dependence in the interaction is counted as the energy dependence for the matrix element given by Eq.\eqref{V elements}, the value of compositeness should deviate from unity. We find the value of deuteron compositeness of this model to be 1.479, which is grater than one. This implies that the energy dependence of the effective interaction which corrects the deuteron binding energy by attractive shift from the constant interaction provides a negative contribution for the value of deuteron elementariness. We find that this problem cannot be resolved by adding higher $q^2$ corrections, $d$-wave contribution nor a more realistic one-pion exchange potential as shown in Table \ref{table_result}. The details of the calculation is shown in Appendix \ref{append.1}.

\begin{center}
	\begin{table}
		\begin{tabular}{ c c c}
			{Model} & {$\breve{X}$} & {Binding Energy/$MeV$} \\
			$c_0$  & 1.000 &  1.477 \\
			$c_0+c_1 q^2$ &  1.479 & 2.232 \\
			$c_0 + c_1 q^2 + c_2 q^4$ & 1.727 & 2.423 \\
			$c_0 + OPEP $ & 1.870 & 2.200(fixed) \\
			$c_0 + c_1 q^2 + OPEP$ & 1.427 & 2.200(fixed)\\
			D channel with $c_0$ & 1.133 & 2.200(fixed) \\
		\end{tabular}
		\caption{Compositeness of deuteron in simple models calculated by Eq.\eqref{X_g} together with the binding energy of deuteron. OPEP stands for one-pion exchange potential. The details of the models are discussed in Appendix \ref{append.1}.}
		\label{table_result}
	\end{table}
\end{center}

\subsection{Perturbation Calculation}
In this subsection, we generalize the discussion given in the previous subsection by proving that energy-dependent attractive correction in the interaction is likely to always provide a value of compositeness lager than unity if one uses Eq.\eqref{X_g}. 

Let us consider a two-body scattering system. We assume that with a constant interaction $v_0$, the system has a bound state with mass $M_0$, and consequently the $T$-matrix of this system has a pole at $E=M_0$. The value of the compositeness for this bound state is unity owing to the constant interaction; $\breve{X}_0 = 1$. Then we consider an attractive energy dependent correction $\tilde v(E)$ in this system. With this correction the interaction is given by $v(E) = v_0 + \tilde v (E)$ and the bound state energy is shifted to $M$. The $T$-matrix of the new system has a pole at $E=M$. These two systems have the same free Hamiltonian $H_0$ and consequently the same loop function $G$. The compositeness of the bound state $E = M$, $\breve{X}$, can be calculated by using Eq.\eqref{X_g}. The details of the calculation are shown in Appendix \ref{append.2}. Finally we find the compositeness of the bound state $M$ in the first order perturbation of $\tilde v(E)$ as

\begin{equation}
	\breve{X}\simeq \breve{X}_0 (1-\lambda \frac{\tilde{v}'}{G'v_0^2}), \label{perturbation}
\end{equation}
which relates compositeness $\breve{X}$ to the compositeness of the unperturbed bound state $\breve{X}_0=1$. Here $\lambda$ is the perturbation parameter indicating order, and $\tilde v^\prime$ and $G^\prime$ are defined as their derivatives evaluated at $E=M_0$
\begin{equation}
	\tilde v^\prime \equiv \left. \frac{\partial \tilde v}{\partial E} \right|_{E = M_0}, \qquad 
	G^\prime \equiv \left. \frac{\partial G}{\partial E} \right|_{E = M_0},
\end{equation}
respectively. 

As discussed in the previous section, the derivative of the loop function $G^\prime$ is negative below the threshold. (We consider bound states, and $M_0$ is smaller than the threshold energy.) Therefore, any positive $\tilde v^\prime$ will have a positive impact on compositeness. Here we consider an attractive energy dependent correction $\tilde v(E)$ which shifts the binding energy $M_0$ to $M<M_0$. In such a situation, $\tilde v^\prime$ should have a positive value considering its definition and expansion at low binding energy regime. It is possible, indeed, that $\tilde v(E)$ with $\tilde v^\prime < 0$ provides a bound state at $E=M<M_0$. Nevertheless, in such a case there exists at least one more bound state at $E>M_0$ and this is out of our consideration because the presence of a bound state higher than $M_0$ is regarded as repulsive correction. In this way, in order to deepen the bound state, it is highly likely to increase compositeness from unity for energy dependent potentials. This implies that the value of elementariness of this state can be negative. This  contradicts with the positivity of elementariness when all of the energy dependence in the integrated interaction stems from the reduction of states and scattering channels. 

This outcome shows that if the calculation based on near-threshold limit gives a value of compositeness larger than unity, it is an implication that the energy dependence cannot be totally attributed to other states and channels, but being inherently in the interaction instead. We need to note that this calculation is based solely on the Lippmann-Schwinger equation under a contact interaction, which means that this calculation did not explicitly take the model space into account, which can be the limitation of the calculation with Eqs.\eqref{X_g} and \eqref{Z_g}.

\subsection{Discussion}
Recall that in subsection \ref{properties of compositeness}, we have discussed the property that other states will bring about positive elementariness based on Schrodinger equation and Feshbach procedure, and we have an expectation to have $0\le \breve{Z}\le1$ as it is to $Z$. However, as shown in this section, if we accept surjective interpretation and the calculation based on it in Eqs.\eqref{X_g} and \eqref{Z_g}, the value of deuteron elementariness $\breve{Z}$ is highly likely to be negative due to the insufficient attraction caused by the scattering length. As a result, we give an evidence that under some circumstances, considering $\breve{X}$ and $\breve{Z}$ as the originally defined $X$ and $Z$ may not be physically appropriate.

\section{Alternative approaches to compositness}
\label{section4}
In this section, the origin of the problem of using $\breve{X}$ in place of $X$ will be discussed. Based on this discussion, some alternative phenomenological approaches will be considered. 

\subsection{Problems on Using \texorpdfstring{$\breve{X}$}{breveX} and Implication to Phenomenology}
To calculate compositeness, a phenomenological potential is required. The path of getting this potential is soley to reproduce experimental observables, and the theoretical considerations of the Hilbert space is skipped, which ought to give a positive elementariness.

Let us look at the one-body state as an example. We notice that the restriction on the sign of the elementary state to be reduced in Eq.\eqref{veff1} is not manifested in the final outcome of  compositeness \eqref{X_calculate} and elementariness \eqref{Z}. From a phenomenological point of view, there might be no limitation to stop the potential from taking a form like
\begin{equation}
	v^{\text{int}}(E)=v-\dfrac{(g_{(0)})^2}{E-M_{B_0}}.
	\label{veff_negative}
\end{equation}
This potential can be interpreted as a negative norm state, and would lead to a negative elementariness. Further, the generalization of this statement from one-body state to two-body and three-body state could be straight-forward. This would show a possibility of finding a negative elementariness in some particles from phenomenology based on surjective interpretation, and deuteron is one example.

In this way, we give a possible reason of why compositeness of deuteron is usually calculated as largen than one, which is taking surjective interpretation and using $\breve{Z}$ interchangably with $Z$.

\subsection{Beyond Effective Range Model}
In a local Hamiltonian QFT under the Effective Range Model, it is found that as long as the effective range is positive, the price must be paid that the space of quantum states will include negative-norm states \cite{Braaten:2007nq}. We can find the same scenario in deuteron, meaning if a calculation on deuteron is to be carried out in this model, negative-norm states will appear for the same reason. A phenomenologcial calculation on NN scattering successfully reproduced the phase shift by including a negative norm state \cite{Kaplan:1996nv}. 

This outcome aligns with our calculation in Eq.\eqref{perturbation} where the effective range will be the dominant contributor of $\tilde{v}'$, making it also the main reason for negative elementariness.

Even though negative norm state may be a reasonable price to pay when only the experimental values are considered, it may not be a good option in the calculation compositeness. Compositeness is not an observable, but rather a value that encapsulates how the particle of interest is to be interpreted. Accepting negative norm states will make it neccessary to find a way to interprete negative elementariness outside the current probability framework of interpretation, weakening the motivation of compositeness as the probability.

On the other hand, while effective range model model is closely related to weak-binding limit and our calculation, the outcomes above may be an indication that effective range model is not appropriate in certain cases.
\subsection{Beyond Surjective Interpretation}

\subsubsection{Loss of Model-Independency}
We may disregard surjective interpretation and accept that energy dependence is intrinsic inside the interaction as
\begin{equation}
	v^{\text{int}}(E)=v(E)+\dfrac{(g_{(0)})^2}{E-M_{B_0}}.
\end{equation}
As we have seen before, substituting the wave function \eqref{comp2_c} into Schrodinger Equation \eqref{comp2_eq2}, we have,
\begin{equation}
	\braket{\psi_0|\psi} = \frac{c g_0}{M_B - M_{B_0}},
\end{equation}
where $c$ originates from the normalization constant when solving the Schrodinger equation. Thus, the elementariness is obtained as
\begin{align}
	Z&=\braket{\psi|\psi_0}\braket{\psi_0|\psi} \notag\\&= - |c|^2 [G\dfrac{d(v^{\text{int}}-v)}{dE}G]_{E=M_B}\notag\\
	&\neq - |g|^2 [G\dfrac{dv^{\text{int}}}{dE}G]_{E=M_B} = \breve{Z}.
\end{align}
From the second line to the third line, we are supposed to use the derivative of Eq.\eqref{veff1}. However, contrary to Eq.\eqref{Z}, in which we have constant $v$, we are unable to ignore the derivative of $v(E)$, and thus the relation $c=g$ breaks. As a result, we are unable to define compositeness and elementariness as they are in Eqs. \eqref{X_g} and \eqref{Z_g}, if there is energy dependence inherent inside the potential instead of interpreted as integrated potential from other channels.

We need to note that even if $c\neq g$, the idendity in Eq.\eqref{identity_g} is not influenced, since it is totally an outcome of Lippmann-Schwinger equation and pole structure.
\subsubsection{Generalization on Interpretation}
We have found that it is theoretically difficult to acquire deuteron property and reasonable compositeness at the same time under surjective interpretation. This is because there should be intrinsic energy dependence in the integrated interaction which cannot be interpreted as the reduction of other explicit states and channels. Thus, we ought to extend the theory to treat energy dependency more thoroughly.

To do this, we will go back to the formalism of compositeness. Specifically, we may compare completeness relations \eqref{identity} and \eqref{identity_g}, which are the identities originated from Schrodinger equation and Lippmann-Schwinger equation. Under surjective interpretation, if we do model calculation, we actually perform the calculation of $X$ and $Z$ twice, once from Schrodinger equation and once from Lippmann-Schwinger equation. We confirm, instead of utilizing, the completeness of Hilbert space. This redundancy aimed at the connection between Schrodinger equation and Lippmann-Schwinger equation, and finally $c=g$. As this method may not be appropriate, we propose to cut off this connection first.

Considering the nature of the Hilbert space and the discussion about positivity of elementariness in Section.\ref{properties of compositeness}, we conclude that compositeness and elementariness both are positive. With the positive sign of the residue $|g|^2$ and normalization factor $|c|^2$, we have
\begin{align}
	[\dfrac{dG}{dE}]_{E=M_B}<0,\\
	[G\dfrac{dv^{\text{int}}-v}{dE}G]_{E=M_B}<0.
\end{align}

With this idea, we can try to recover a similar form from Eq.\eqref{identity_g}
\begin{align}
	&\acute{X} = - |g|^2 [\dfrac{dG}{dE}]_{E=M_B} = \breve{X},\\
	&\acute{Y} = - |g|^2 [G\dfrac{dv}{dE}G]_{E=M_B},\\
	&\acute{Z} = - |g|^2 [G\dfrac{d(v^{\text{int}}-v)}{dE}G]_{E=M_B} ,
\end{align}
where we defined a new quantity $\acute{Y}$, which we call as interactioness. The new set of quantities are labeled with an acute mark to separate from both the original formalism and surjective interpretation. Because of Eq.\eqref{identity_g}, $\acute{X}$ and $\acute{Z}$ are both positive, and $\acute{X} + \acute{Y} + \acute{Z} = 1$. Therefore, we can interprete the interactioness $\acute{Y}$ as a part of compositeness, so that the compositeness can be less than one.

The reason for this set of quantities to be chosen is that it bears strong physical relation with the original formalism. Even though the equivalence between $c$ and $g$ is broken, we can view Eq.\eqref{gc_renormalization} as a generalization from equivalence to relation, where the model independence is lost due to the explicit existence of $v$. Finally, this set of quantities can smoothly connect to surjective interpretation with $\dfrac{dv}{dE}=0$.

From here, we can generalize the word "interpretation" as a way to separate $v$ from $v^{\text{int}}$, or $\acute{Y}$ and $\acute{Z}$, or equivalently how the model space is viewed. From this view point, surjective interpretation is one among the infinite number of interpretations, and is a special case where no intrinsic energy dependent potential exists. As a simple example, we can define another extreme case of interpretation that all the energy dependence of $v^{\text{int}}$ is intrinsic, i.e. originates from $v$. Under this interpretation, any bound states origin from scattering states and always have elementariness zero, i.e. totally composite particle.

To obtain a clearer understanding of interpretation, let us assume that we have exact knowledge of $T$ matrix. This is not practically possible, but can make the consequences straighforward. From the knowledge of a channel of interest, we are able to acquire its loop function $G$, and from Lippmann-Schwinger equation \eqref{LSeq_eff}, $v^{\text{int}}$ can be solved. By implementing some interpretation, we get to know $v$ since the interpretation limits the intrinsic energy dependence. No further information or assumptions are required in the calculation of compositeness and elementariness. With exact knowledge of the experiment, fixing interpretation may lead to reasonable compositeness and elementariness.

\subsection{Beyond Quantum Mechanics}

In the construction of calculation of compositeness, we work in the momentum space and required that the interaction is always local. It is in most of the cases a reasonable choice, but the interaction does not always have to take such form. For example, there are other formalisms that work in the coordinate space.\cite{Aoki2020} This effect may be important in deuteron because pion exchange plays an important role. Such diagram shall form a ladder diagram where the off-shell part of pion can be important, and a low energy expansion in the momentum space may not be able to catch such characteristics.  In Ref.\cite{Savage:2003ud}, a pionless effective theory that contains an energy dependent interaction is shown, and this interaction provides a positive effective range. This interaction plays a similar role as some negative norm state in the integrated interaction even though such state does not exist in the original theory. The energy dependent interactions in the pionless effective theory may be obtained from pionful theories by integrating out the pion degrees of freedom.

We propose to modify the formalism of compositeness completely, preferably with field theory approach. In QFT, we have well defined models such as Chiral Perturbation Theory that can eliminate the arbitrariness in interactions. On the other hand, it is relatively difficult to interprete compositeness, because compositeness relies on states, and states are not the main entities discussed in field theory. Some work has been done in this direction. \cite{Guo2016,Oller2018}
\section{Conclusion}
\label{section:conclusion}
In this paper, we have formalized compositeness and elementariness in the case that the interaction contains an intrinsic energy dependence. We then discussed the separation of Schrodinger equation and Lippmann-Schwinger equation driven by the energy dependence.

Since none of the numerical calculations can reproduce a reasonable compositeness for deuteron under current assumption, we have performed a perturbative calculation on compositeness. We have found that if we follow surjective interpretation strictly, we can have a direct relationship between binding energy and compositeness: if the binding energy indicated by scattering length is lower than that from experiment, the compositeness will likely always be larger than one, and vice versa. On the other hand, we concluded that the elementariness induced by other states should be positive. Through proof by contradiction, we have concluded that surjective interpretation is not appropriate for deuteron. The origin of such outcome can be traced back to energy dependence in the interaction that cannot be expressed by other states or channels.

In order to solve this problem, we proposed a  straightforward extension by defining a new quantity interactioness, and thus the generalized interpretation can be discussed case by case. However, this extension will require an exact knowledge on the interaction. Alternatively, we propose to modify the theory from quantum mechanics into field theory formalism. Even though there may be some difficulty in interpretation since states are not prominent in QFT, the interactions are more well-defined.

At the current state, it may be possible to utilize compositeness as a way to benchmark effective theories. Since the positivity of elementariness is proved in this paper, such positivity should hold when we take an interpretation under the framework of an effective theory.

\section*{Acknowledgement}
We would like to show our gratitude to Prof. Tetsuo Hyodo, Prof. Makoto Oka, Dr. Kotaro Murakami for the discussions and insights during this research. The work of Z.Y. was supported by Japanese Science and Technology Agency (JST), the establishment of university fellowships towards the creation of science technology innovation, Grant Number JPMJFS2112. The work of D.J. was partly supported by Grants-in-Aid for Scientific Research from JSPS (JP21K03530, JP22H04917 and JP23K03427).
\appendix

\section{Alternative Formalism without Explicit Energy Dependence}
\label{append.0}
In this section, we will give an alternative formalism that assumes no explicit energy dependence but only velocity dependence in the form of $\bs{V}(q^2,q'^2,E(q^2))$ and show these two formalisms will give equivalent calculation of compositeness. To achieve this, an ansatz of cutting off by order in an low-energy approximation fashion is required, and this ansatz can be validated by the fact that it can recover the bound state condition as in Lippmann-Schwinger equation. 

We can assume a form of interaction that resembles the separable form as
\begin{equation}
\braket{{q}_j|\boldsymbol{V}|{q'}_k}=v_{jk}(E_q)f_j(q^2)f_k(q'^2),
\end{equation}
where $E_q$ is the energy fixed from momentum $q$ through the on-shell condition \eqref{onshell}, meaning that the state $\ket{q}$ is on the mass shell as required by old-fashioned perturbation theory (OFPT)\cite{Schwartz:2014sze}. We find that the energy of intermediate states are not in the potential $v_{jk}$. This is because the interaction $v_{jk}$ is already complete even without a dependency on $E_{q'}$ as long as we absorb any of the energy dependence in form factor $f_k(q'^2)$.

Similar to subsection \ref{subsection.singlechannel}, we will discuss the one scattering channel and one elementary state case as demonstration. The coupled-channel Schrodinger equations can be written in a similar fashion,
\begin{align}
	\braket{q|\bs{H}|\psi} &= f(q^2)\int \dfrac{d^3q'}{(2\pi)^3} v(E_q) f(q'^2)\tilde{\psi}(q') \notag\\
	&+E_q\tilde{\psi}(q)+g_{(0)}f(q^2)\braket{\psi_0|\psi}= M_B\tilde{\psi}(q),
\end{align}
\begin{align}
	\braket{\psi_0|\bs{H}|\psi}=M_{B_0}\braket{\psi_0|\psi}+g_{(0)}\int &\dfrac{d^3q'}{(2\pi)^3}f(q'^2)\tilde{\psi}(q') \notag\\
	& =  M_B\braket{\psi_0|\psi}.
\end{align}
By eliminating $\braket{\psi_0|\psi}$ from the equation, we have a new form for the integrated potential,
\begin{align}
	E_q\tilde{\psi}(q)+f(q^2)v^{\text{int}}(E_q,M_B)\int \dfrac{d^3q'}{(2\pi)^3}&f(q'^2)\tilde{\psi}(q')\notag\\
	&= M_B \tilde{\psi}_j(q),
	\label{append.0.sch_eq}
\end{align}
\begin{equation}
	v^{\text{int}}(E_1,E_2)=v_\alpha(E_1,E_2) + v_\beta(E_2),
\end{equation}
where
\begin{align}
	v_\alpha(E_1,E_2) &= v(E_1)-v(E_2),\\
	v_\beta(E_2) &= v(E_2) + \dfrac{(g_{(0)})^2}{E_2-M_{B_0}}.
\end{align}

Similar to the original formalism, the formal solution of Eq.\eqref{append.0.sch_eq} can be written as
\begin{align}
	\tilde{\psi}(q) = - \dfrac{[c_\alpha(E_q) + c_\beta]f(q^2)}{E_{q}-M_B},
	\label{append.0.wavefunction}
\end{align}
where,
\begin{align}
	c_\alpha(E_q)&=v_{\alpha}(Eq, M_B)\int \dfrac{d^3q'}{(2\pi)^3}f(q'^2)\tilde{\psi}(q'), \label{c_alpha}\\
	c_\beta&=v_{\beta}(M_B)\int \dfrac{d^3q'}{(2\pi)^3}f(q'^2)\tilde{\psi}(q'). \label{c_beta}
\end{align}
We would like to note that $c_\alpha(M_B)=0$, which will be used several times in this formalism.

Observe the integral part,
\begin{align}
	h \equiv \int \dfrac{d^3q'}{(2\pi)^3}f(q'^2)\tilde{\psi}(q') = \int \dfrac{d^3q'}{(2\pi)^3}\dfrac{[c_\alpha(E_{q'}) + c_\beta][f(q'^2)]^2}{M_B-E_{q'}},
\end{align}
we find that $c_\alpha(E_{q'})$ will have a higher order of $M_B-E_{q'}$ as long as $v$ follows Taylor expansion around $M_B$, so that $h$ is domiated by $c_\beta$ instead of $c_\alpha$. As a result, we make an ansatz that ignores the $c_\alpha$ term,
\begin{align}
	h \simeq \int \dfrac{d^3q'}{(2\pi)^3}\dfrac{c_\beta[f(q'^2)]^2}{M_B-E_{q'}}.
	\label{append.0.h}
\end{align}

By putting Eq.\eqref{append.0.h} back to Eq.\eqref{c_beta}, we can recover the bound state condition of Lippmann-Schwinger equation in the same way as the original formalism,
\begin{equation}
	1-v^{\text{int}}(M_B)\int \dfrac{d^3q'}{(2\pi)^3}\dfrac{[f(q'^2)]^2}{M_B-E_{q'}} = 0,
\end{equation}
so that this ansatz is verified, and the Green's function $G$ will also take the same form,
\begin{equation}
	G(E) = \int \dfrac{d^3q}{(2\pi)^3}\dfrac{[f(q^2)]^2}{E-E_{q}},
	\label{append.0.G}
\end{equation}

Under the same ansatz, we can calculate compositeness with Eqs.\eqref{append.0.wavefunction}\eqref{append.0.G}, 
\begin{align}
	X &= \int \dfrac{d^3q}{(2\pi)^3}|\tilde{\psi}(q)|^2 \simeq  \int \dfrac{d^3q}{(2\pi)^3}\dfrac{|c_\beta|^2 [f(q^2)]^2}{(E_{q}-M_B)^2} \notag\\
	&= -|c_\beta|^2[\dfrac{dG}{dE}]_{E=M_B}\notag \\
	&= -[|c|^2]_{E_q=M_B}[\dfrac{dG}{dE}]_{E=M_B}.
\end{align}

Similarly, we can calculate the elementariness
\begin{align}
	Z&=\braket{\psi|\psi_0}\braket{\psi_0|\psi} \notag\\
	&\simeq |c_\beta|^2 G(M_B)\dfrac{g_{(0)}g_{(0)}}{(M_B-M_{B_0})^2}G(M_B)\notag\\
	&= - |c_\beta|^2 [G\dfrac{d(v^{\text{int}}-v)}{dE}G]_{E=M_B}\notag\\
	&= -  [|c|^2]_{E_q=M_B}[G\dfrac{d(v^{\text{int}}-v)}{dE}G]_{E=M_B}.
\end{align}

We can find that the form of $X$ and $Z$ are recovered with a slight difference between $|c|^2$ in the explicit energy dependent formalism and $[|c|^2]_{E=M_B}$ in the alternative formalism. This difference is caused by the fact that $c$ is originally a constant in explicit energy dependent formalism while a variable in the alternative one. However, we argue that the difference is not important considering the physical meaning of $c$ as the renormalization factor and the low energy nature of such formalisms. Further, under surjective interpretation, $c$ is inferred from $g$ which is the residue of the pole, so that it is only practically meaningful at pole position.

\section{Calculation on Deuteron under Surjective Interpretation}
\label{append.1}

In this section, we will calculate compositeness with surjective interpretation as an application of our theoretical consideration. To calculate the compositeness of deuteron, we  take the binding energy of deuteron as 2.2MeV, and effective range expansion coefficients $a=5.424$ fm, $r_0=1.759$ fm, $r_1 = 0.040 \text{fm}^3$ \cite{Babenko2007}. Even though we have much more accurate values on deuteron binding energy, we take 2.2MeV for simplicity because it will not influence the analysis since we are trying to account for the problem of deuteron compositeness above one instead of its exact value. We consider a model with isospin symmetry and nucleon mass is defined to be $938.9185$ MeV. The calculation is performed with Lippmann-Schwinger equation.

We choose the loop function to be a relativistic dimensional regularized loop function \cite{Hyodo2012}, with the form of

\begin{align}
	G &= -i (4 M^2) \int \frac{d^4 q}{(2\pi)^4} \frac{1}{(P-q)^2-M^2+i\epsilon} \frac{1}{q^2-M^2+i\epsilon} \notag \\
	&= \frac{4 M^2}{16\pi^2}\{a_\Lambda + \frac{q}{E} ( 2ln(E^2+2qE)-2ln(-E^2+2qE)) \},
\end{align}
which is practically equivalent to a non-relativistic one given by Ref.\cite{BEANE2001}
\begin{equation}
G = - \frac{iqM}{4\pi},
\end{equation}
in this energy region due to the consequence of optical theorem. The substraction constant $a_\Lambda$ is determined through a mapping with 3-momentum cutoff scheme at threshold

\begin{equation}
	a_\Lambda = -\frac{M}{2\pi^2}\Lambda,
\end{equation}
We choose the cutoff to be $150$ MeV. We need to note that the loop function in 3-momentum cutoff scheme is non-relativistic since a relativistic one has to be described by elliptic functions. However, we expect no difference between them at threshold where we perform the comparision.
\subsection{Low Energy Expansion}
The first model we consider is low energy expansion model. To get the low energy constants, we perform a mapping of S-matrix between the theoretical potential and experimental effective range expansion:

\begin{equation}
	S=1-\frac{qM}{2\pi}\frac{1}{\frac{1}{v^{\text{int}}}-G}=1+\frac{2iq}{q cot\delta -iq},
\end{equation}
in which
\begin{align}
	v^{\text{int}} &= c_0 + c_1 q^2 + ... ,\\
	qcot\delta &= -\frac{1}{a} + \frac{1}{2}r_0q^2+r_1q^4+... ,
\end{align}
from the effective range expansion.

We can first look at the outcome of the 0th order, namely the constant interaction. At this order, as long as the sign of the interaction is correctly chosen, a pole will always exist. Due to the mathematical structure of Eq.\eqref{X_g}, this pole is always of compositeness one. By fixing this constant to the scattering length from the experiment, the binding energy is $1.477$ MeV compared with $2.2$ MeV from the experiment value. This pole has a lower binding energy, but with a reasonable compositeness.

Since the attraction is not enough for reasonable deuteron energy at the 0th order, the 1st order correction needs to be attractive, which aligns the 1st order interaction factor with effective range. By adding this order, the binding energy is modified to $2.232$ MeV, which is close to the observation. However, with the improvement in the binding energy, the compositeness turns out to be 1.479. This outcome can be interpreted by considering a theory with an explicit deuteron pole. An explicit deuteron pole should be related with positive elementariness, but the effective range and deuteron energy indicate a pole structure with negative residue, which will result in negative elementariness, and thus compositeness higher than one.
\subsection{One Pion Exchange Potential}
We then made a calculation on one pion exchange potential(OPEP), which is normally written as
\begin{align}
	V_\pi(r) =\frac{m_\pi^2}{12\pi} (\frac{g_A}{\sqrt{2}f_\pi})^2(\tau_1\cdot\tau_2)[S_{12} (1+\frac{3}{m_\pi r} +\frac{3}{(m_\pi r)^2})\notag\\+\sigma_1\cdot\sigma_2]
	\frac{e^{-m_\pi r}}{r} - \frac{4\pi}{3} \sigma_1\cdot\sigma_2\delta^3(r).
\end{align}
in coordinate space. In this equation, $g_A$ and $f_\pi$ are constants about interaction, $m_\pi$ is pion mass, $\tau$ is Pauli matrix representing isospin, $\sigma$ is Pauli matrix for spin, and $S_{12}$ is the tensor operator. Besides, in the case of deuteron, we have already known that $S=1$ and $T=0$, thus $\sigma_1\cdot\sigma_2 = 1, \tau_1\cdot\tau_2 = -3, S_{12}=0$. Additionally, the term with $\delta^3(r)$ in S-wave will finally result in a constant in momentum space, and can be absorbed in the low energy constant from the 0th order of this expansion. Simplifying the OPEP potential into S-wave with the discussion above, we have
\begin{align}
	V_\pi(r) =c_{opep}
	\frac{e^{-m_\pi r}}{r},
\end{align}
in which $c_{opep} = -\frac{m_\pi^2}{4\pi}(\frac{g_A}{\sqrt{2}f_\pi})^2$.
Performing Fourier transformation, we have
\begin{align}
	V_\pi (q_\pi)= c_{opep} \frac{4\pi}{m_\pi}\frac{1}{q_\pi^2+m_\pi^2}.
\end{align}
Projecting it into S wave componet, since S wave dominates in deuteron, we have
\begin{align}
	V_\pi(q) = c_{opep}\frac{\pi}{2q^2} ln \frac{4q^2 +m_\pi^2}{m_\pi^2},
\end{align}
where $q$ is the center of mass momentum of two nucleon system. Since $c_{opep}$ is negative, this interaction is always attractive, and can result in a higher binding energy. By combining it with the constant term, we will get compositeness $X=1.870$. 

We need to note that the potential in this subsection is not strictly a one pion exchange potential in a field theoretical sense. It is because we projected it into the S wave and the potential expresses propagation of a massless pion. On the other hand, the pion exchange should finally generate pion loops as we can see from Eq.\eqref{LSeq_eff}. The pion momentums ought to be integrated through, and the case calculated here is a low momentum approximation instead of an on-shell approximation from field theoretical point of view. This may make the interaction non-realistic, but will not influence the discussion of this paper.
\subsection{Yamaguchi Potential}
To further fix the problem in the previous subsection, we consider a finite range interaction with Yamaguchi Potential, which is defined as a modification of form factor as
\begin{equation}
	f(q) = \frac{\beta^2}{q^2+\beta^2},
\end{equation}
in which, $q$ is center of mass frame momentum, and $\beta$ is an energy scale related with finite range interaction.

By fixing the scattering length and the effective range to observation in a $V=c_0+c_1 q^2$ form, we find that compositeness increases as binding energy increases. The compositeness will become 1 at $\beta = 336 MeV$ with binding energy $B=2.01 MeV$. Further increasing $\beta$ will result in a higher than 1 compositeness according to the calculation above. Since we are aiming at a finite range interaction, we will include further terms $V=c_0+c_1 q^2 + c_2 q^4$. To reproduce the binding energy, the compositeness turns to be $X = 1.21$. Thus, finite range interaction cannot be a reason of unreasonable compositeness.
\subsection{Coupled Channels}
We then go to two channel coupling case, at next to next to leading order (NNLO), the interaction is given by Ref.\cite{Epelbaum2006}
\begin{equation}
	V=\begin{pmatrix}
		c_0+c_1 q^2+c_2 q ^4 & c_{SD,1} q^2 + c_{SD,2} q^4\\
		c_{SD,1} q^2 + c_{SD,2} q^4& c_{DD} q^4
	\end{pmatrix}.
\end{equation}
Since we do not expect any elementariness in crossing channels, i.e. no energy dependence should exist in crossing channels, we transform the $T$ matrix, $G$ matrix and potential through

\begin{align}
	\overset{*}{T} &= P^{-1} T P^{-1},\\
	\overset{*}{V} &= P^{-1} T P^{-1},\\
	\overset{*}{G} &= P G P,
\end{align}
in which matrix $P$ is given by
\begin{equation}
	P=\begin{pmatrix}
		1 & 0\\
		0& q^2
	\end{pmatrix}.
\end{equation}

We have more parameters than we need, but when the pole gets deeper, the compositeness in the $S$ channel will get higher for any change parameter whenever we reproduce the deuteron binding energy, the scattering length, and the effective range. Examples of the outcome are shown in the Table \ref{table_result}, but we cannot achieve reasonable compositeness in a 2 channel Low Energy Constant (LEC) model up to Next-to-next-to leading order (NNLO).

\subsection{Square Well Potential}
This potential is defined as
\begin{equation}
	V(r) = \begin{cases} 
		A & r < R \\
		0& r > R 
	\end{cases}.
\end{equation}
The fourier transformation of it is
\begin{equation}
	V(q) = \frac{4\pi A}{q} (-\frac{1}{q} R cos(Rq) + \frac{1}{q^2} sin(Rq)).
\end{equation}
The parameters are fixed as $R = 1$ fm. If we fix the scattering length to experimental value, we will have binding energy $B=1.472$ MeV and compositeness $X = 1.046$. By changing the parameter of this potential, the 0th order expansion which is related with the scattering length is also influenced. We further include a core on top of a well to reproduce the repulsive core of a nucleon-nucleon interaction. We take $r_{core} = 1$ fm, and fix the scattering length from experiment. The core is repulsive and the $R$ can be set to be effective in a larger range. We have a free parameter to move, but we failed to reproduce reasonable deuteron properties and positive elementariness at the same time.
\subsection{Discussion on Outcome of Numerical Calculation}
We find that as long as the scattering length, or scattering property at threshold, is fixed, all of these models shares the same behavior that any attempt to deepen the deuteron pole will result in a higer compositeness than 1, if one uses Eq.\eqref{X_g}. This outcome aligns with the outcome from weak binding limit which is 1.67. Besides, the expansion is made with regard to momentum, and higher-order terms are expected to have much less influence on the state. Further, any model will have a projection on this LEC model, which suggests that any model, as long as it follows an analytical expansion around threshold, should not be able to reproduce deuteron pole with a reasonable compositeness.

\section{Theoretical calculation on compositeness from Lippmann-Schwinger Equation}
\label{append.2}

\subsubsection{Unperturbed Calculation as Preparation}
\label{generalized calculation}
Assuming $f(q^2)=1$ for simplicity, we consider two systems with V-matrix elements $v_0$ and $v$, which deviate by a infinitesimal $\tilde{v}$. $v_0$ is energy independent, while $\tilde{v}$ can be energy dependent:
\begin{equation}
	v(E)=v_0+\tilde{v}(E).
\end{equation}
These systems share the same free Hamiltonian $H_0$ and the T-matrix elements are obtained as
\begin{align}
	t_0 &= \frac{1}{1-v_0G(E)}v_0,\\
	t &= \frac{1}{1-v(E)G(E)}v(E),
\end{align}
where $G(E)$ is the Green's function of $H_0$. Each system has a bound state with mass $M_0$ and $M$respectively. The $T$-matrix elements accordingly have a pole at $E=M_0$ and $E = M$, which can be expressed as
\begin{align}
	t_0&\equiv\frac{N_0(E)}{E-M_0}\label{N0},\\
	t&\equiv\frac{N(E)}{E-M},
\end{align}
in which we defined function $N_0$ and $N$ as the numerators.

The difference between the two systems comes from $\tilde{v}$, with a different pole position and residue. We define self-energy $\Sigma(E)$ by
\begin{equation}
	t = \frac{N_0(E)}{E-M_0+\Sigma(E)},
	\label{tN0}
\end{equation}
Thus, we have
\begin{align}
	N(E) &= N_0(E) \frac{E-M}{E-M_0+\Sigma(E)}\notag\\ &= N_0(E)\frac{E-M_0+\Sigma(M)}{E-M_0+\Sigma(E)}.
	\label{NN0}
\end{align}
Since the numerator and denominator of Eq.\eqref{NN0} are both zero at $E\rightarrow M$, we use L'Hospital's rule and obtain
\begin{equation}
	N(M) = N_0(M) \frac{1}{1+\dot{\Sigma}}. \label{NatM}
\end{equation}
As a convention, for any function $F$ we define 
\begin{align}
	F'&\equiv \frac{\partial F}{\partial E}|_{E=M_0},\\
	\dot{F} &\equiv \frac{\partial F}{\partial E}|_{E=M},
\end{align}
where higher order derivatives follow the same fashion without being explicited defined here. From Eqs.\eqref{N0}, \eqref{tN0} and Lippmann-Schwinger equations, we find
\begin{align}
	E-M_0 &= N_0(E)(\frac{1}{v_0}-G(E)),\\
	E-M_0+\Sigma(E) &= N_0(E)(\frac{1}{v}-G(E)),
\end{align}
and thus the self-energy is given as
\begin{align}
	\Sigma(E) &= N_0(E)(\frac{1}{v_0+\tilde{v}(E)}-\frac{1}{v_0}).
	\label{SigmaE_unperturbed}
\end{align}
For later use, we calculate $N_0(M_0)$ and its derivative with Eq.\eqref{N0},
\begin{align}
	N_0(M_0) &= \lim_{E\rightarrow M_0}\frac{E-M_0}{\frac{1}{v_0}-G}=-\frac{1}{G'}\label{N0M0},\\
	\frac{\partial N_0(E)}{\partial E} &= \frac{ (\frac{1}{v_0} - G) - (E-M_0)(-\frac{\partial G}{\partial E}) }{(\frac{1}{v_0} - G)^2}\notag\\
	&=\frac{1+N_0(E)\frac{\partial G}{\partial E}}{\frac{1}{v_0} - G}.
	\label{B13}
\end{align}
We find that both the numerator and denominator of Eq.\eqref{B13} are zero at $E=M_0$, and we use L'Hospital's rule to evaluate it as
\begin{equation}
	N_0'= \lim_{E\rightarrow M_0} \frac{
		\frac{\partial N_0(E)}{\partial E} \frac{\partial G}{\partial E}
		+
		N_0(E) \frac{\partial^2 G}{\partial E^2}
	}{ -\frac{\partial G}{\partial E}}
	=-N_0'-N_0(M_0)\frac{G''}{G'} .
\end{equation}
By connecting the left and right hand side, we get
\begin{equation}
	N_0' = -\frac{1}{2}N_0(M_0)\frac{G''}{G'}.
	\label{N0'}
\end{equation}
\subsubsection{Perturbative Calculation}
We are calculating the properties of the bound states. We propose a modification to the potential as $v=v_0+\lambda \tilde{v}$ with a perturbative order factor $\lambda$, original potential $v_0$ and modifed potential $\tilde{v}$. The energy and compositeness are also modified with the change in potential as
\begin{align}
	M &= M_0 +\lambda M_1 +\lambda^2 M_2 +... \quad ,\\
	\breve{X} &= \breve{X}_0 +\lambda \breve{X}_1 +\lambda^2 \breve{X}_2+ ... \quad .
\end{align}
The goal is to find a general relationship between the modified potential $\tilde{v}$ and the first order perturbation of compositeness $\breve{X}_1$. For a small modification in potential, we can expect the first order perturbation term to prevail.

The calculation will be performed with a constant $v_0$, which means we start from a constant interaction theory. From the definition of compositeness and elementariness, we can see that a constant interaction will always result in compositeness one, i.e. $\breve{X}_0=1$. Besides, the constant term is solely determined by scattering length. These factors make such interaction a good starting point. 

First, we consider the energy modification $M_1$. The energy is defined as a pole position of $T$ matrix,
\begin{align}
	v_0G_0(M_0) &= 1,\label{V0G0=1}\\
	v(M)G(M)&=1.
	\label{VG=1}
\end{align}
By expanding Eq.\eqref{VG=1} with respect to $\lambda$, and utilize Eq.\eqref{V0G0=1}, we obtain,
\begin{align}
	M_1 &= -\frac{\tilde{v}(M_0)}{G'v_0^2},\\
	M_2 &= (\frac{\tilde{v}'}{\tilde{v}(M_0)} + G'v_0 -\frac{G''}{2G'} )M_1^2.
\end{align}
From Eq.\eqref{SigmaE_unperturbed}, we have
\begin{align}
	\Sigma(E) &\simeq -\lambda N_0(E)\frac{\tilde{v}(E)}{v_0^2},
	\label{SigmaE_perturbed}
\end{align}
up to the first order of $\lambda$. Compositeness is calculated by Eq.\eqref{X_g} as 
\begin{align}
	\breve{X} = - N(M)\dot{G},
\end{align}
and with Eq.\eqref{NatM}, $\breve{X}$ up to the first order of $\lambda$ is obtained as
\begin{align}
	\breve{X} \simeq - N_0(M)\frac{1}{1+\dot{\Sigma}}G'(1+\lambda \frac{G''}{G'} M_1).
	\label{Xhalf}
\end{align}
There are two parts yet to be calculated, $N_0(M)$ and $\dot{\Sigma}$. From Eq.\eqref{N0}, we obtain
\begin{align}
	N_0(M) &= \frac{M-M_0}{\frac{1}{v_0}-\frac{1}{v(M)}}\notag\\
	&\simeq N_0(M_0)[1-\frac{\lambda}{2}\frac{G''}{G'}M_1].
	\label{N0M}
\end{align}
On the other hand, for $\frac{1}{1+\dot{\Sigma}}$, with Eq.\eqref{SigmaE_perturbed}, Eq.\eqref{N0M0}, and Eq.\eqref{N0'},
\begin{align}
	\frac{1}{1+\dot{\Sigma}}&\simeq 1+ \lambda \frac{N_0'\tilde{v}(M_0)+N_0(M_0)\tilde{v}'}{v_0^2}\notag\\
	&= 1-\lambda(\frac{1}{2}\frac{G''M_1}{G'}+\frac{\tilde{v}'}{G'v_0^2}).
	\label{fSigma}
\end{align}
Combining Eqs.\eqref{Xhalf}, \eqref{N0M}, \eqref{fSigma}, we can calculate the compositeness up to first order as
\begin{align}
	\breve{X}&\simeq -N_0(M_0)[1-\frac{\lambda}{2}\frac{G''}{G'}M_1]\notag\\
	&\times [1-\lambda(\frac{1}{2}\frac{G''M_1}{G'}+\frac{\tilde{v}'}{G'v_0^2})]G'(1+\lambda \frac{G''}{G'} M_1).
\end{align}
The outcome is
\begin{equation}
	\breve{X}\simeq \breve{X}_0 (1-\lambda \frac{\tilde{v}'}{G'v_0^2}).
\end{equation}

\bibliography{bib}

\end{document}